\documentstyle[aps,prd,floats]{revtex}
\newcommand{\bba}{\begin{eqnarray}}
\newcommand{\eea}{\end{eqnarray}}
\newcommand{\bb}{\begin{equation}}
\newcommand{\ee}{\end{equation}}
\newcommand{\bban}{\begin{eqnarray*}}
\newcommand{\eean}{\end{eqnarray*}}

\def\a{\alpha}
\def\b{\beta}

\def\d{\delta}

\def\f{\phi}

\def\r{\rho}

\def\t{\tau}

\def\G{\Gamma}

\preprint{DAMTP-2000-97, Brown-HET-1235}
\date{8 September 2000}
\tighten

\begin{document}
\title{Cosmological consequences of the brane/bulk interaction}

\author{C. van de Bruck${}^1$\thanks{Electronic address: C.VanDeBruck\,@\,damtp.cam.ac.uk},
M. Dorca${}^2$\thanks{Electronic address: dorca\,@\,het.brown.edu},
C.J.A.P. Martins${}^{1,3}$\thanks{Electronic address: C.J.A.P.Martins\,@\,damtp.cam.ac.uk},
M. Parry$^4$\thanks{Electronic address: m.parry\,@\,ic.ac.uk}}

\address{${}^1$ Department of Applied Mathematics and Theoretical Physics, 
Center for Mathematical Sciences,\\
University of Cambridge, Wilberforce Road, Cambridge CB3 0WA, U.K.}

\address{${}^2$ Department of Physics, Box 1843, Brown University, Providence, RI 02912, USA}

\address{${}^3$ Centro de Astrof\'{\i}sica, Universidade do Porto,
Rua das Estrelas s/n, 4150-762 Porto, Portugal}

\address{${}^4$ Blackett Laboratory, Imperial College, Prince Consort Road, London SW7 2BZ, U.K.}

\noindent
\maketitle
\begin{abstract}
We investigate cosmological consequences arising from the 
interaction between a homogeneous and isotropic brane--universe
and the bulk. A Friedmann equation is derived which incorporates
both the brane and bulk matter contributions,
which are both assumed to be of arbitrary fluid form.
In particular, new terms arise which describe 
the energy flow onto (or away from) the brane, as well as changes of the 
equation of state in the bulk. We discuss Randall--Sundrum type 
models as well as dilatonic domain walls and carefully consider 
the conditions for stabilizing the induced gravitational 
constant. Furthermore, consequences for cosmological perturbations are 
analyzed. We show that, in general, super-horizon amplitudes 
are not constant. 
\end{abstract}
\pacs{PACS numbers: 04.50.+h, 11.15.Mj, 12.10.-g, 98.80.Cq}
\pacs{Keywords: gravitation, Einstein field equations, membrane
theory, perturbation theory, cosmology}

\section{Introduction}
The idea that the universe is a brane embedded in a higher dimensional
space has attracted a lot of attention in the last two years
\cite{branecos}---for earlier proposals see \cite{early}. 
It was found that the Friedmann equation 
on the brane contains corrections to the usual four--dimensional
equation. In particular a term $H\propto \rho$ was found
\cite{Binetruy}. This term is observationally problematic
({\em e.g.} for nucleosynthesis, among other things), 
but the model is consistent if one considers a cosmological 
constant in the bulk and the tension on the brane, which leads 
to a cosmological version of the Randall--Sundrum scenario \cite{RS}
of warped geometries \cite{RSCOS}.

On the other hand, particle 
physics models predict the existence of scalar fields, both on the 
brane and in the bulk. In fact, in heterotic M--theory, the 11--dimensional 
strong coupling limit of the E$_8 \times$E$_8$ heterotic string
theory, one particular bulk scalar field measures the deformation 
of the Calabi--Yau space and can not arbitrarily set to be zero.
It is actually one of the aims of string cosmology
to investigate the consequences of the evolution of these moduli
fields. Two interesting examples are: i) in brane world scenarios, scalar
fields are thought to stabilize the fifth dimension (see
{\em e.g.} \cite{csaki}); ii) the energy of bulk scalar field(s) 
might flow onto or away from the brane universe and thereby modify 
the expansion dynamics. 

But cosmology is more than the evolution of the background. One of the
biggest puzzles in cosmology is the origin and evolution of the
structures we see in our universe. In the last twenty years it became
a working hypothesis to search for the origin of
these structures in the physics 
of the micro-cosmos. The most popular paradigm is the inflationary
universe \cite{inflainfo}. Here one speculates that the seeds of
the structures are quantum fluctuations in the inflaton field which
are stretched onto macroscopic length scales during a `super-luminal'
expansion epoch (commonly called inflation), where 
they become classical. Inflation leads to definitive observational
signals, {\em e.g.} in the form of peaks and valleys in the spectrum
of microwave background anisotropies and there is very good hope to
check and severely constrain the parameter
space of inflationary models with 
ongoing balloon experiments and future satellite experiments. In the 
context of brane world scenarios it is crucial to investigate if new 
effects appear and how they influence the evolution of perturbations. 
This may then provide a way to test these theories. 

At the time of writing, brane world cosmology is still in its infancy.
Issues of inflationary model building in this context have just begun
to be considered, and so far the evolution of density fluctuations
has been relatively neglected in these scenarios.
Recently, investigations started 
in this direction, see \cite{us} and \cite{pertur}. Such a 
formalism allows for an investigation of the full 5D field equation 
for perturbations of first order. 

In this paper we turn our attention to the interaction between 
the bulk and the brane, which is another non-trivial aspect of brane 
world theories. In particular, we will discuss the flow 
of energy onto or away from the brane--universe. The presentation
is as follows: in  section II we discuss 
the evolution of the background. We derive a Friedmann equation on 
the brane, which includes a term describing the energy flow from the fifth 
dimension. In section III we discuss the evolution of perturbations,
and in particular we derive the equation
of motion for the gauge invariant curvature variable 
$\zeta$. As we will show, this variable is not conserved even for
adiabatic perturbations, in contrast to the ``usual''
four--dimensional world, where $\zeta$ is conserved. 

\section{Background evolution of the brane universe}
In this paper we will consider only one brane, which describes our
universe, embedded in a five--dimensional space. In particular, we
are going to follow  the formalism and notations introduced in \cite{us}.
The field equation is (we set the five-dimensional gravitational
constant $\kappa_5 \equiv1$),

\begin{equation}\label{fieldeq}
G_{\a\b}\equiv R_{\a\b} - \frac{1}{2}g_{\a\b} R = T_{\a\b} 
+ T_{\a\b}^{(b)} \delta(y-y_b).
\end{equation}
where we consider a metric with the following form
\begin{equation}\label{unmetric}
ds^2 = a^2 b^2 \left( dt^2 - dy^2 \right) - a^2 \Omega_{ij} dx^i dx^j,
\end{equation}
with $y$ denoting the coordinate along the fifth dimension, 
and $\Omega_{ij}$ the metric of a 3--dimensional space of constant
curvature, that is,

\bb
\Omega_{ij}=\d_{ij}\left[1+\frac{k}{4}x^lx^m\d_{lm}\right]^{-2},
\ee
where $k=0,1,-1$ corresponds to a flat, closed or open universe,
respectively. Also we assume that the two
scale factors depend on both $t$ and $y$, i.e. $a=a(t,y)$ and $b=b(t,y)$.
Note that we have chosen a conformal gauge for the $t$-$y$ 
part of the metric. In this gauge the brane is static and located at 
$y=y_b$, which we assume to be a {\em fix point} of a $Z_2$--symmetry along
the fifth dimension  (see \cite{us} for details). 

We shall make the mild assumption that the matter both on the bulk
and brane is describable by a fluid. Thus, the 
energy--momentum tensor for the bulk matter is
\bb
 {T^\a}_\b = \left(\begin{array}{ccc}
             \r_B&0&-r_B\\0&-p_B{\d^i}_j&0\\r_B&0&-q_B
             \end{array}\right)\; .
 \label{T}
\ee
where $r_B$ represents the energy flow onto the brane. This term is
non-vanishing, for example, when there is a scalar field on the bulk
which depends on both coordinates $t$ and $y$. 
The brane matter is described by
\bb
{T^{(b)~\a}}_\b = \left(\begin{array}{ccc}\r^{}&0&0\\
                  0&-p^{}{\d^i}_j&0\\
                  0&0&0\end{array}\right)\; .
 \label{Tn}
\ee
For the sake of clarity, we shall denote bulk quantities with a subindex $B$,
while quantities without subindex are brane quantities. Then,
Einstein's equations for the metric (\ref{unmetric}) can be found
to be

\bba
 a^2b^2{G^0}_0 &\equiv& 3\left[2\frac{\dot{a}^2}{a^2}+\frac{\dot{a}\dot{b}}{ab}
                   -\frac{a''}{a}+\frac{a' b'}{ab}+kb^2\right]
                = a^2b^2\left[\r_B +\r^{ }
                \bar{\d}(y-y_b)\right]\label{G00}\\
 a^2b^2{G^5}_5 &\equiv&3\left[\frac{\ddot{a}}{a}-\frac{\dot{a}\dot{b}}{ab}
                 -2\frac{{a'}^2}{a^2}-\frac{a' b'}{ab}+kb^2\right]
                 =-a^2b^2q_B \label{G55}\\
 a^2b^2{G^0}_5&\equiv&3\left[ -\frac{\dot{a}'}{a}+2\frac{\dot{a}a'}{a^2}
                 +\frac{\dot{a}b'}{ab}+\frac{a'\dot{b}}{ab}\right]
                 = -a^2b^2r_B \label{G05}\\
 a^2b^2{G^i}_j&\equiv&\left[ 3\frac{\ddot{a}}{a}+\frac{\ddot{b}}{b}-
                 \frac{\dot{b}^2}{b^2}-3\frac{a''}{a}-\frac{b''}{b}
                 +\frac{{b'}^2}{b^2}+kb^2\right]{\d^i}_j
                 =-a^2b^2\left[p_B+p^{ }
                 \bar{\d}(y-y_b)\right]{\d^i}_j ,\label{Gij}
\eea
where the delta-function $\bar\d$ incorporates a factor $1/ab$, and we
use the notation ${\dot f}=\partial f/\partial t$,
${ f'}=\partial f/\partial y$, for a given function $f$. 
Using standard methods, it is easy to derive the jump conditions for
the metric components. We are going to assume here that our brane
corresponds to the brane located at $y=R$ in ref.~\cite{us}.
These jump conditions are simply (see \cite{us} for details)

\bb\label{[a,b]}
{a'\over a}= {1\over 6}ab\r^{ }\; ,\qquad
{b'\over b}= -\frac{1}{2}ab\left(\rho^{ }+p^{ }\right)\; .
\ee
Restricting the $05$ component~(\ref{G05}) to the brane and 
using the jump conditions above, we can easily find

\bb\label{conservB}
\dot{\r}^{ }=-3{{\dot a}\over a}\left(\r^{ } +
                  p^{ }\right)+ 2abr_B\; .
\ee
This is the energy conservation equation on the brane,
which has been  generalized to allow a possible 
energy flow $r_B$ onto the brane. 
The sign of $r_B$, according to (\ref{T}), will determine
if the energy flow is {\it onto} the brane or away from
it. Equivalently, the restriction of the  $55$ component to the
brane gives,
 
\bb\label{eins}
 \frac{\ddot{a}}{a}-\frac{\dot{a}\dot{b}}{ab}+kb^2
  = -{a^2b^2\over 3}\left[\frac{1}{12}\r^{ }\left(\r^{ }+3p^{ }\right)
    +q_B\right]\; .
\ee

From these relations one can proceed to derive the Friedmann
equation, {\em i.e.} to find a first integral of (\ref{eins}). 
We start by re-parametrising the time--coordinate on the 
brane as follows 

\bb\label{branet}
d\t=(ab)dt .
\ee
Thus, we change to the {\em proper time} on the brane
(recall that on the brane the scale factors $a$ and $b$ are functions
of $\t$ alone). Then, using the notation
$f_\t=df/d\t$ for a given function $f$, equation (\ref{eins}) changes
to

\bb\label{zwei}
 \frac{{a}_{\t\t}}{a} + \left(\frac{{a}_\t}{a}\right)^2 + \frac{k}{a^2}
  = -{1 \over 3}\left[\frac{1}{12}\r^{ }\left(\r^{ }+3p^{ }\right)
    +q_B\right]\; ,
\ee
whereas the energy--conservation equation (\ref{conservB}) now reads

\bb\label{conservB2}
\frac{d{\r}^{ }}{d\t}=-3{{a_\t}\over a}\left(\r^{ } +
                  p^{ }\right)+ 2r_B\; .
\ee
Following \cite{flana}, the easiest way to proceed is
to write the scale factor $a$ as 
$a = \exp \alpha$. Also, we use $H={\dot a}/a$ as the standard notation
for the {\em Hubble parameter}, and for convenience
we define another Hubble parameter referred to the proper time on the
brane (\ref{branet}) as ${\cal H}=a_\t/a$. Then, we easily get the
following relations,

\begin{equation}
\frac{{a}_{\t\t}}{a} + \left(\frac{{a}_\t}{a}\right)^2 = \alpha_{\t\t} 
+ 2 \dot \alpha_\t^2, 
\end{equation}
and 
\begin{equation}
2  \alpha_{\t\t} = \frac{d {\cal H}^2}{d \alpha},
\end{equation}
so that one gets
\begin{equation}\label{derH}
\frac{d({\cal H}^2 e^{4\alpha})}{d \alpha} = - 
e^{4\alpha} \left[ \frac{\rho(\rho + 3p)}{18} + \frac{2q_B}{3}\right] 
-\frac{2k}{e^{2\alpha}}e^{4\alpha}.
\end{equation}
If equation (\ref{conservB2}) is now used to write the pressure $p$ 
as a function of $\rho$ and $d\rho/d\t$ and is then inserted in equation
(\ref{derH}) one can easily derive
\begin{equation}\label{last}
\frac{d({\cal H}^2 e^{4\alpha})}{d \alpha} = 
\frac{d}{d\alpha} \left[ e^{4\alpha} \frac{\rho^2}{36} \right]
- \frac{e^{4\alpha}}{9}\frac{r_B\rho}{\cal H} - \frac{2}{3}e^{4\alpha} q_B
-\frac{d}{d\alpha} \left(e^{4\alpha}\frac{k}{e^{2\alpha}}\right).
\end{equation}
Then, if we define two functions ${\cal Q}$ and ${\cal E}$ by
\begin{eqnarray}
\frac{d}{d\alpha} e^{4\alpha}{\cal Q} = e^{4\alpha} q_B, \label{qs} \\ 
\frac{d}{d\alpha} e^{4\alpha}{\cal E} = e^{4\alpha}
\frac{r_B\rho}{\cal H} 
\label{rs},
\end{eqnarray}
we find the solution of eq. (\ref{last}) to be
\begin{equation}\label{fried}
{\cal H}^2 = \frac{\rho^2}{36} -\frac{2}{3} {\cal Q} - \frac{1}{9}{\cal E} 
- \frac{k}{a^2} + \frac{{\cal A}}{a^4},
\end{equation}
where ${\cal A}$ is an integration constant. 
Finally, it is useful to change to ordinary time derivatives in
equations (\ref{qs}) and (\ref{rs}). This gives
\begin{eqnarray}
\frac{d{\cal Q}}{d\t} + 4{\cal H}{\cal Q} &=& {\cal H}q_B\; , \label{Q}\\
\frac{d{\cal E}}{d\t} + 4{\cal H}{\cal E} &=& r_B\rho \label{E} \;.
\end{eqnarray} 
Equations (\ref{fried})--(\ref{E}) (together with (\ref{conservB2})) 
are the evolution equations for the system we discuss. Furthermore, 
they allow a comparison to other existing approaches. Note that, 
for a scalar field in the bulk, the set of equations (\ref{fried}), 
(\ref{Q}) and (\ref{E}) is equivalent to equation (12) in 
\cite{soda}, written here in physical variables. 

Inspired by heterotic M--theory or the Randall--Sundrum model, we may write 
$\rho = \rho_M + U_B$, where $\rho_M$ is the ``ordinary'' brane matter 
(which by definition  does not couple to the bulk matter) and
$U_B$ is a contribution from the (projected) bulk matter. It then follows
that

\begin{equation}\label{calH1} 
{\cal H}^2 = \frac{\rho_M ^2}{36} + \frac{1}{18}U_B \rho_M 
+ \frac{U_B^2}{36} - \frac{2}{3}{\cal Q} - \frac{1}{9}{\cal E}
- \frac{k}{a^2} + \frac{{\cal A}}{a^4} \; . 
\end{equation}
This equation has a similar form as in the Randall--Sundrum (RS)
case. Indeed, to get the Friedmann equation, we just need to identify

\begin{equation}\label{gravconst}
48 \pi G = U_B.
\end{equation}
Furthermore, the split we have chosen ($\rho = \rho_M + U_B$) 
suggests that both matter components evolve separately. In fact, by 
definition, $U_B$ alone describes the interactions with the bulk. Hence
the energy--conservation equation (\ref{conservB2}) splits into

\begin{eqnarray}
\frac{dU_B}{d\t} &=& -3{\cal H}U_B\left(1 + w_U (\t) \right)+2r_B \label{cons1}\;
, \label{e1}\\ 
\frac{d\rho_M}{d\t} &=& -3{\cal H}\left(\rho_M + p_M \right) \label{cons2}.
\end{eqnarray}
Here, $w_U$ may be seen as the effective equation of state of the 
(projected) bulk matter. 
Note that this provides a specific example of a model where a
four-dimensional ``effective'' constant can evolve due to dynamical
effects in higher dimensions. This fact
alone may be a powerful tool
in constraining brane--world models (see \cite{cmb} and references therein).

For the particular case of $r_B=0$ and $q_B=const=\rho_B$ 
(which represents the RS--case), 
we easily get from equation (\ref{calH1}) and the definitions of 
${\cal Q}$ and ${\cal E}$ 

\begin{equation}
{\cal H}^2 = \frac{\rho_M ^2}{36} + \frac{8 \pi G}{3} \rho_M 
+ \Lambda_{\rm eff} + \frac{{\cal C}}{a^4} - \frac{k}{a^2}, 
\end{equation}
which is in fact the equation found in the literature. Here, we have 
defined an effective cosmological term
\begin{equation}
\Lambda_{\rm eff} = \frac{1}{6}\left( \frac{U_B^2}{6} - q_B \right).
\end{equation}
Note that, in general, the integration constant ${\cal C}$ is 
different from ${\cal A}$.

Some comments are in order at this point.
In the model presented so far, the
evolution of the induced gravitational constant $G$ (\ref{gravconst})
is {\it universal}, if $U_B$ (and thus $\rho_B$) is not stabilized,
i.e. if $dU_B/d\t\neq 0$. However, we have two possibilities to
stabilize $G$:

\noindent
i) At late times, the equation of state of the projected bulk matter is 
$w_U = -1$ and there is no energy flow, that is $r_B=0$. 
This is in fact the case in Randall--Sundrum type models, in which, 
for example, the bulk is of anti--de Sitter type. Note that such a model 
allows for an early evolution of $G$, because it is conceivable 
that the anti--de Sitter space might be a natural endpoint of the 
evolution of the bulk. 

\noindent
ii) At late times (at least), the energy flow {\em ``tracks''} the 
expansion of the brane universe, that is

\begin{equation}
r_B = \frac{3}{2}{\cal H} U_B\left( 1+w_U (\t) \right)\; .
\end{equation}
In such a model, it is imaginable that the bulk space reacts against
changes of the dynamics on the brane, for example if the effective 
equation of state on the brane changes.

The first option has been discussed extensively in the literature.
The second  option, however, allows for several new insights 
in the stabilization of $G$ in
the framework of brane world theories. In particular, 
note that the dynamics of the
five--dimensional bulk is contained in the quantity $r_B$, 
therefore the system of
equations (\ref{fried})-(\ref{E}) and (\ref{cons1})-(\ref{cons2}),
which describe the dynamics on the branes, 
is {\it not} closed.

As a particular case, we are going to assume that we have
a scalar field in the
bulk. Such a situation arises, for example, in five--dimensional 
heterotic M--theory \cite{hetor}, or in dilatonic brane worlds 
\cite{dilbrane} (see also \cite{mazum}). We are going to denote the
bulk--scalar field by $\phi$, its potential energy
in the bulk by $U$ and its projected energy on the brane by $U_B$. 
From the energy--momentum tensor of the bulk matter 
(see \cite{hetor} for details) one finds 

\begin{equation}\label{rscalar}
r_B = -T^{0}_{~5} = -\frac{\phi_\t}{4}\frac{\partial U_B}
{\partial \phi}\; ,
\end{equation}
and
\begin{equation}\label{zwischen}
q_B = -T^{5}_{~5} = -\frac{1}{4}\phi_\t^2 
    + \frac{1}{16}\left( \frac{\partial U_B}{\partial \phi}\right)^2 
-\frac{1}{2}U
\end{equation}
In the expressions of $r_B$ and $q_B$ we have already used the boundary
condition for the bulk scalar field on the brane. This condition, which can
be found in ref.~\cite{hetor}, is
 
\begin{equation}
\phi' = \frac{ab}{2} \frac{\partial U_B}{\partial \phi}.
\end{equation}
From these expressions we can construct an effective cosmological 
constant in the Friedmann equation as follows. According to
\cite{dilbrane}, and in order to define $\Lambda_{\rm eff}$, we collect only 
those terms in the Friedmann equation which contain 
$U$, $U_B$ and the derivative $\partial U_{B} / \partial \phi$.
Since the quantities $U$ and $U_B$ are slow time-varying for any 
physically motivated model, then  we get from equation (\ref{Q})

\begin{equation}\label{Lambda}
\Lambda_{\rm eff} = \frac{1}{6} \left[ \frac{U_B^2}{6} 
+ \frac{1}{2}U -\frac{1}{16}\left( \frac{\partial U_B}{\partial\phi} \right)^2 \right].
\end{equation}
The information that one can extract from this exercise is
that the effective cosmological
constant, as well as the induced gravitational constant $G$ on the
brane, are influenced by the $05$-- and the $55$--component of the 
bulk energy--momentum tensor. In particular, the evolution of the induced 
gravitational constant
is described by eqns. (\ref{gravconst}), (\ref{cons1}) and (\ref{rscalar}).
Note for instance that expression (\ref{Lambda}) for the cosmological 
constant is in agreement with \cite{dilbrane}. 

Apart from the effective cosmological constant, the Friedmann equation
on the brane contains an integral over $\phi_\t$, coming from equation
(\ref{Q}), and also a radiation term. Furthermore, the differential
equation for ${\cal E}$ (\ref{E}) can be solved 
(see \cite{dilbrane} for details). Thus,
in \cite{dilbrane} different solutions of dilatonic brane worlds where
found. This is a path that we do not intend to pursue here. 

To summarize, we have written the Friedmann equation in terms of 
$q_B$ and $r_B$, which respectively are a time--integral 
of the $55$--component of the bulk stress--energy 
tensor and a time--integral over the energy--flow onto the
brane. Such a situation differs from the derivation in
ref.~\cite{dilbrane} on the fact that in their case
the bulk--dependence enters also through the projection of the 
five--dimensional Weyl--tensor on the brane. 

\section{Cosmological perturbations and the brane/bulk interaction}

In the discussion so far we have restricted ourselves to the background
evolution. However,
as we will see in this section, interesting consequences appear also 
when considering cosmological perturbations. Our goal is to
derive the equations of motion for the curvature 
perturbation in a rather general setup. The basic reason we have
in mind to do this is the following. 
It was shown in \cite{malik} 
that the curvature of slices of constant density is conserved whenever
energy conservation holds. Now, as discussed throughout the previous section,
energy conservation in brane worlds is not warranted and an energy flow 
onto or away from the branes has important consequences. Given the results of
\cite{malik}, it is only natural to expect a modification of the 
evolution for super-horizon perturbations. 

To discuss the evolution of density perturbations, we will choose a
particular gauge  to simplify the calculations.
In \cite{us} we have shown that in five dimensions 
we can use a ``generalized'' longitudinal gauge. We shall use the same
gauge choice here. It defines a unique coordinate system and
the advantages have been recognized in the literature. An important 
difference from the Randall--Sundrum gauge is that the perturbation 
of the 55--component is not zero. However, the perturbation variables
coincide in this gauge with the gauge--invariant variables and thus 
we have control over gauge artifacts.

In what follows we will discuss scalar perturbations only. 
Moreover, we will limit ourselves to vanishing anisotropic stress (along
the slices $t=y=$constant) in the bulk as well as on the brane, which 
is the case for scalar fields. This assumption further ensures that the
use of the ``generalized longitudinal gauge'' is appropriate. 

With the assumptions listed above, the perturbed metric takes the 
form \cite{us}
\begin{eqnarray}
ds^2 &=& a^2b^2 \left[(1+2\phi)dt^2 - (1-2\Gamma) dy^2  - 2 W dydt
              \right]-a^2\Omega_{ij}(1 - 2\psi)dx^i dx^j. 
\end{eqnarray}

The perturbed energy-momentum tensor has the form 
\begin{equation}
\d {T^a}_\b = \left(\begin{array}{ccc} \d\r_B&-(\r_B
    +p_B)b^{-2}v_{B|j}&-
\d r_B\\
              (\r_B +p_B)v_B^{|i}&-\d p_B\,{\d^i}_j&-u_B^{|i}\\
              \d r_B+2r_B(\f +\G )-(\r_B +q_B)W&-b^{-2}u_{B|j}&
              -\d q_B\end{array}\right)
\end{equation}
for the bulk matter and
\begin{equation}
\d {T^{a}}_\b = \left(\begin{array}{ccc}
       \d\r&-(\r+p)b^{-2}v^{ }_{|j}&-\d r^{}\\
       (\r+p)v^{ |i}&-\d p{\d^i}_j&0\\
       \d r^{}-\r^{ }W&0&0\end{array}\right)
 \label{dTn}
\end{equation}
for the brane matter. 

The energy conservation for the brane--perturbations to first order 
can be found to be
\begin{eqnarray}\label{conservation}
(\delta \rho)^{.} &=& - 3 H \left(\delta \rho + \delta p\right) 
                    + 3\left(\rho + p\right)\dot\psi 
                    \nonumber \\
                    &-& 2a^2\left(\rho 
                    + p\right)\left(\rho_B + q_B\right)v
                    +2ab\left(\Gamma + 2\phi + \frac{\delta r_B}{r_B}\right)r_B,
\end{eqnarray}
where we have neglected all spatial derivatives because we are interested
in super-horizon amplitudes. All quantities are evaluated {\it on} the 
brane. This equation can be obtained by taking the jump of the
05--component of Einstein's equation across the position of the brane
(see \cite{us} for details). 

Eqn. (\ref{conservation}) has the same form as in four dimensions,
apart from the last two terms which describe the energy--flow 
onto or away from the brane. Of course, according to the
equation above, the total energy  (brane+bulk) is conserved.  
However, due to the inhomogeneities on the brane, the 5$D$ geometry
is distorted in such a way that gravitational energy from the extra 
dimension can flow onto the brane. In fact,
such an statement is valid for any brane world scenario. Also,
according to the setup we have used,
equation  (\ref{conservation}) is  general. Finally, recall that 
if the bulk is in its vacuum state (i.e. $\rho_B + q_B = 0$ and
$r_B=0$) and also $\d r_B=0$, then the usual 4$D$ conservation law is valid.

Usually in the 4$D$ case the matter perturbation is related to the 
metric perturbation via a gauge invariant variable $\zeta$, defined as
\cite{bar83}
\begin{equation}\label{zeta}
\zeta = -\psi - H \frac{\delta \rho}{\dot \rho}.
\end{equation}
This term makes sense in $5D$ cosmology under the assumptions that
it is restricted on the brane and it transforms according to
the induced $4D$ metric on the brane. Then, it is easy 
to calculate 
\begin{equation}\label{dotzeta}
\dot \zeta = -\dot\psi - H\frac{(\delta \rho)^.}{\dot \rho} 
             - 3 H^2(1+c_s^2)\frac{\delta \rho}{\dot \rho} -
             \frac{\delta \rho}{\dot \rho ^2}\left[ 4abr_B \dot H 
             -2(abr_BH)^.\right],
\end{equation}
where we have used the energy conservation for the background
(\ref{conservB}) and the definition $c_s^2 = \dot p/\dot \rho$.
Using the following definition for the pressure perturbation

\begin{equation}
\delta p = c_s^2 \delta \rho + \delta p_{{\rm nad}},
\end{equation}
where $\delta p_{\rm nad}$ is the non--adiabatic part of the pressure 
perturbation on the brane (describing the entropy production), we get 
from eqns. (\ref{conservation}) and (\ref{dotzeta}) 

\begin{eqnarray}\label{main}
\dot \zeta = - \dot\psi \left[ 1 - \frac{1}{1-F} \right] 
             - \frac{H}{1-F} \frac{\delta p_{{\rm nad}}}{\rho + p} 
             - \frac{1}{1-F} \left[ \frac{2}{3} a^2 v(\rho_B +q_B) \right] \nonumber\\ 
             + \frac{2}{3(1-F)}\frac{ab}{\rho+p}\left(\Gamma + 2\phi +
               \frac{\delta
             r_B}{r_B}\right)r_B
             - \frac{\delta \rho}{\dot \rho^2}\left[ 4abr_B \dot H 
             -2(abr_BH)^.\right],
\end{eqnarray}
with $F = 2abr_B/3H(\rho+p)$. 

There is important information coming from (\ref{main}). Namely,
as long as the extra dimension is not in a vacuum state
(that is, $\rho_{B} + q_B \neq 0$
and/or $r_B\neq0$), {\it the super-horizon amplitudes on the brane 
are not necessarily
constant, even if the perturbations are adiabatic}. An amplification
of perturbations was previously thought to happen at the end of
inflation (preheating) \cite{preheat}. In that case the second term 
in equation (\ref{main}) would be important. However,
in the case of brane worlds 
the amplification depends also on the history of the extra dimension.
Thus, it is important to find out how and when the extra dimension becomes 
stabilized, as this can have obvious observational consequences.

In the case of the often discussed Randall--Sundrum world, the last
three terms in (\ref{main}) vanish (see also the discussion in 
\cite{tatsi}). More generally speaking, the bulk space has to 
settle into a vacuum state in the very early universe, otherwise 
the amplitude of adiabatic scalar perturbations will not remain 
constant. Models without this property will have extreme 
difficulties in surviving observational tests.
However, for the particular case where the cosmological constant 
in the bulk is provided 
by a bulk scalar field (i.e. $r_B=0$ and $\rho_B + q_B= 0$), 
the perturbation of the energy flow onto (or away from) the 
brane will be present (i.e. the $\delta r_B$--term). 
In that case, the evolution 
of a adiabatic perturbations on super-horizon scales will be described
by
\begin{eqnarray}
\dot \zeta = \frac{2}{3}\frac{ab\delta r_B}{\rho+p}.
\end{eqnarray}
This equation tells us that the perturbation of the energy flow 
will modify super-horizon perturbations on large scales, and provides a
clear observational handle on these models.

Whereas our argumentation is based on the form of the
energy--conservation on the brane it was shown recently by Gordon 
and Maartens that other bulk effects produce non--adiabatic modes on the 
brane, which implies an effective growth of super-horizon
perturbations \cite{gordon}. All these results make inflationary
model building, among other things, considerably more restrictive
than in the usual four--dimensional models. 

\section{Discussion}

In this paper we have investigated the effects of the interaction 
between a brane universe and the bulk in which it is embedded. 
We have argued that in the case where the bulk scalar field
contributes to the dynamics on the brane, the induced
four--dimensional gravitational constant depends on the 
energy flux onto (or away) from the brane as well as on changes 
of the equation of state in the bulk. 

Whenever a scalar field is in the bulk, an energy flow onto (or way
from) the brane universe is possible.
In these models, the dynamics of the scalar
field has to be studied in detail, in order to make statements about
the gravitational constant in four dimensions. Apart from the
gravitational constant, an energy flux onto the brane changes 
its expansion dynamics. For example, one can even imagine
models in which a cosmological constant and a black hole exist in the
bulk \cite{chamblin}. In that case, the time--dependent mass of the black
hole influences the dynamics on the brane. 

The conclusions we have drawn hold strictly only for a single brane
universe. There will be some changes if one considers more than
one brane. In the case of two branes, 
for example, where energy is globally conserved and propagates between the 
two branes, the four--dimensional gravitational constant will depend 
on the boundary conditions {\it on both branes}. To what extend a 
bulk scalar field influences the dynamic on ``our'' brane world will 
thus depend on the boundary condition on the other brane. But we
expect that some of the statements remain valid in these models in 
the era where the universe is effectively five
dimensional. Furthermore, we have assumed that the five--dimensional 
gravitational constant is time--independent. 

We have also investigated the consequences of energy conservation violation 
for large scale perturbations in brane world models. We have shown
that, in general, adiabatic super-horizon perturbations are not
constant. The amplitude will depend crucially on the orbifold history and 
on the stabilization mechanism. We believe this will provide a powerful
way to test and rule out many realizations of the `brane-world'
paradigm.

\acknowledgements

C.v.d.B. is supported by the Deutsche
Forschungsgemeinschaft (DFG). M. Dorca is supported by
the {\em Fundaci\'on Ram\'on Areces}. The research was supported
in part (at Brown) by the U.S. Department of Energy under Contract 
DE-FG02-91ER40688, TASK A. C.M. is funded by FCT (Portugal)
under `Programa PRAXIS XXI', grant no. PRAXIS XXI/BPD/11769/97.

\end{document}